\documentclass[prl,twocolumn,graphics]{revtex4}

\usepackage{graphicx}
\usepackage{amssymb}
\usepackage{dcolumn}
\usepackage{bm}
\input epsf
\input{epsf.sty}

\begin{document}


\title{Anomalous quasiparticle transport in the superconducting state of  CeCoIn$_5$ }

\author{Y.~Kasahara$^1$, Y.~Nakajima$^{2}$, K.~Izawa$^{2,3}$, Y.~Matsuda$^{1,2}$, K.~Behnia$^{2,4}$, H.~Shishido$^5$, R.~Settai$^5$, and Y.~Onuki$^5$}
\affiliation{$^1$Department of Physics, Kyoto University, Kyoto 606-8502, Japan}%
\affiliation{$^2$Institute for Solid State Physics, University of Tokyo, Kashiwanoha, Kashiwa, Chiba 277-8581, Japan}%
\affiliation{$^3$CEA-Grenoble, 38054 Grenoble cedex 9 France}%
\affiliation{$^4$Laboratoire de Physique Quantique (CNRS), ESPCI, 10 Rue de Vauquelin, 75231 Paris, France}%
\affiliation{$^5$Graduate School of Science, Osaka University, Toyonaka, Osaka, 560-0043 Japan}%

\date{\today}

\begin{abstract}

We report  on a study of thermal Hall conductivity $\kappa_{xy}$ in the superconducting state of CeCoIn$_5$.  The scaling relation and the density of states of the delocalized quasiparticles, both obtained from $\kappa_{xy}$,  are consistent with $d$-wave superconducting symmetry.  The onset of superconductivity is accompanied by a steep increase in the thermal Hall angle,  pointing to a striking enhancement in the quasiparticle mean free path.  This enhancement is drastically suppressed in a very weak magnetic field.  These results highlight that  CeCoIn$_5$ is unique among superconductors.  A small Fermi energy, a large superconducting gap, a short coherence length,  and a long mean free path all indicate that CeCoIn$_5$ is clearly in the superclean regime ($\varepsilon_F/\Delta\ll\ell/\xi$), in which peculiar vortex state is expected.

\end{abstract}

\pacs{74.20.Rp, 74.25.Bt, 74.25.Fy, 74.70.Tx}

\maketitle

Five years after the discovery of superconductivity in CeCoIn$_5$ \cite{Petrovic}, this compound has become the focus of considerable attention. Indeed, CeCoIn$_5$ occupies a particular place among unconventional superconductors; it shares more features with high-$T_c$ cuprates than any other heavy-fermion (HF) superconductor.  Most importantly,  superconducting instability arises in the normal state that exhibits pronounced non-Fermi-liquid behavior due to the proximity of an antiferromagnetic (AFM) quantum critical point(QCP) \cite{Sidorov}.  Several measurements indicate that the superconducting gap  has $d$-wave symmetry with line nodes perpendicular to the plane \cite{Movshovich,Izawa,Ormeno,Rourke}.

CeCoIn$_5$ exhibits several fascinating properties, which have never been observed in any other superconductor. In a strong magnetic field, the superconducting transition is of the first order, indicating a field-induced destruction of the superconducting state by Pauli paramagnetism \cite{Izawa,Bianchi2}.  Closely related to this,  the emergence of a spatially inhomogeneous Fulde-Ferrel-Larkin-Ovchinnikov superconducting state has been reported in the vicinity of the upper critical field \cite{FFLO,Kakuyanagi}.  Recent NMR spectra also have revealed an unusual electronic structure in the vortex core \cite{Kakuyanagi}.  Moreover the observation of a QCP in the vicinity of the upper critical field $H_{c2}$ for $H\parallel c$ suggests that superconductivity prevails, preventing the development of the AFM order \cite{QCP}.

Another issue of interest is the increase in the quasiparticle (QP) lifetime below $T_c$, indicated by thermal conductivity $\kappa_{xx}$ and microwave  experiments \cite{Movshovich,Izawa,Ormeno}.  This feature of CeCoIn$_5$, reminiscent of very clean high-$T_c$ cuprates, is not observed in other HF superconductors.  Thermal Hall conductivity $\kappa_{xy}$, the non-diagonal element of the thermal conductivity tensor in a perpendicular magnetic field,  is a powerful probe of this feature; it is purely electronic and the direct consequence of a transverse QP current, while $\kappa_{xx}$ includes both electronic and phononic contributions. Over the past few years, the study of the thermal Hall effect in high-$T_c$ cuprates has opened a new window on  QP transport \cite{Zhang,Zeini,Simon,Durst,Ong}.

In this Letter, we report on a study of longitudinal and transverse thermal conductivities of CeCoIn$_5$.  The results highlight a steep increase in the QP mean free path $\ell$ directly inferred from the temperature dependence of the thermal Hall angle $\Theta \equiv \tan^{-1} \kappa_{xy} / \kappa_{xx}$.  The magnitude of $\ell$ estimated in this way can be compared with that extracted from the QP thermal diffusivity (i.e. the ratio of thermal conductivity to specific heat) and confirms the unusually small Fermi energy $\varepsilon_F$ in CeCoIn$_5$.  On the other hand, even a small magnetic  field leads to a dramatic decrease in $\ell$.  This phenomenon, yet to be understood, is unique to CeCoIn$_5$.  We also found that $T$- and $H$-dependence of  $\kappa_{xy}$ supports the  $d$-wave symmetry.

Single crystals of  CeCoIn$_{5}$ ($T_c$~=~2.3~K) were grown by the self-flux method. Both $\kappa_{xx}$ and $\kappa_{xy}$ were measured by the steady-state method by applying the heat current along the [100] direction with  {\boldmath $q$} $\parallel$ {\boldmath $x$} for {\boldmath $H$} $\parallel$  {\boldmath $c$}.  The thermal gradients $-\nabla_x T \parallel$  {\boldmath $x$} and $-\nabla_y T \parallel$
{\boldmath $y$} were measured by  RuO$_2$ thermometers.   
 Above 0.4~K, no hysteresis was observed in sweeping $H$ \cite{measure}.   The sign of $\kappa_{xy}$ is negative,   as for the electrical Hall conductivity $\sigma_{xy}$. 

\begin{figure}[t]
\begin{center}
\includegraphics[height=60mm]{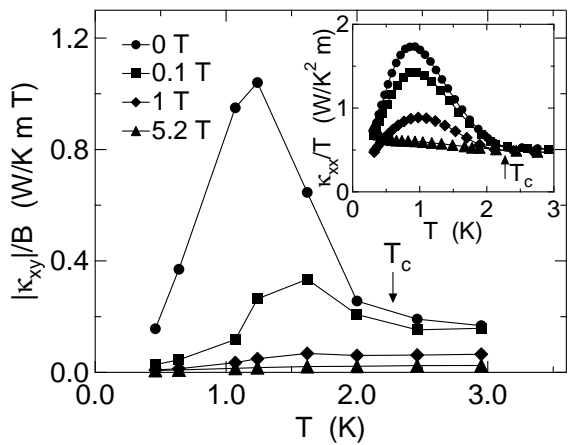}
\caption{ Temperature dependence of the thermal Hall conductivity divided by $B$, $|\kappa_{xy}|/B$.  The zero field limit is obtained from the $H$-linear dependence of $|\kappa_{xy}|$ at low field (see Fig.~3(a)).  Inset: Temperature dependence of  $\kappa_{xx}/T$ in $H$. }
\end{center}
\end{figure}

The inset of Fig.~1 shows the $T$-dependence of $\kappa_{xx} /T$.  In zero field, upon entering the superconducting state, $\kappa_{xx}/T$ display a kink and exhibits a pronounced maximum at $\sim$0.8~K \cite{Movshovich,Izawa}.     Figure 2 (a) depicts the $H$-dependence of $\kappa_{xx}$.   Applying $H$, $\kappa_{xx}$ decreases up to $H_{c2}$ after showing an initial steep decrease.   Figure 2(b) and the inset depict the $H$-dependence of $|\kappa_{xy}|$.   A strong non-linear $H$-dependence is observed in $|\kappa_{xy}|$.    Similar to $\kappa_{xx}$, the absolute slope of $|\kappa_{xy}|$ versus $H$ at high fields is reduced as the temperature is lowered.   
 The transition to the normal state below $\sim$1~K for both $\kappa_{xx}$ and $|\kappa_{xy}|$ is marked by a pronounced jump, indicating a first-order transition \cite{Izawa,Bianchi2}.   (In CeCoIn$_5$ the upper critical field determined by the orbital effect $H_{c2}^{orb}$ is nearly 2.5 times larger than $H_{c2}$; $H_{c2}^{orb}\agt$ 12~T.)

At  low fields,  as shown in the inset of Fig.~2(b), $|\kappa_{xy}|$ exhibits a steep increase with a linear dependence on $H$.   At $T \leq $0.64~K, $|\kappa_{xy}|$ exhibits a prominent peak at $\sim$0.06~T.    It should be noted that a similar peak structure in $\kappa_{xy}$ has also been reported  for ultra-clean YBCO single crystals \cite{Zhang}.  In Fig.~1, we plot  the $T$-dependence of $|\kappa_{xy}|/B$ and the initial Hall slope $|\kappa_{xy}^0|/B\equiv\lim_{B\rightarrow0}|\kappa_{xy}|/B$.   The overall temperature dependence of $|\kappa_{xy}^0|/B$ is similar to $\kappa_{xx}$; as the temperature falls below $T_c$,  it  exhibits a pronounced maximum at $\sim$1~K.    This behavior of $|\kappa_{xy}^0|/B$ again bears a striking resemblance to YBCO \cite{Zhang,Zeini}.

\begin{figure}[t]
\begin{center}
\includegraphics[height=80mm]{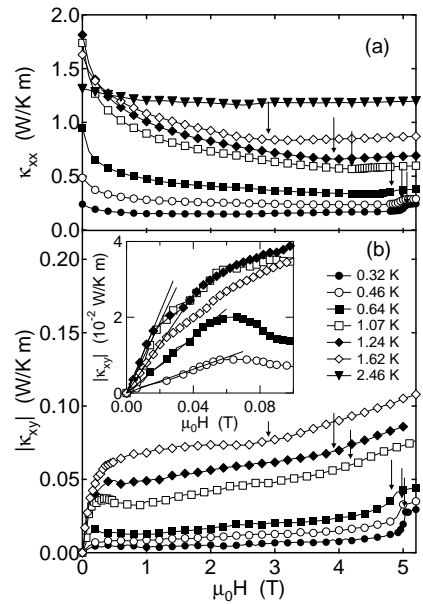}
\caption{ (a) Field dependence of  thermal conductivity $\kappa_{xx}$.  Arrows indicate $H_{c2}$ determined by resistivity measurements.  (b) The same data for the thermal Hall conductivity $|\kappa_{xy}|$.  Inset: Thermal Hall conductivity in the low field regime.  Thin solid lines represent the initial Hall slope.  }
\end{center}
\end{figure}

Before discussing the QP transport, let us examine the validity of the ``transverse''  Wiedemann-Franz (WF) law.  Just above $T_c$, $\kappa_{xy}$ and  $\sigma_{xy}$ yield a ``transverse'' Lorenz number very close to the expected WF value:  $L_{xy}=\lim_{B\rightarrow0}
\kappa_{xy}/\sigma_{xy}T\simeq 1.05L_0$ (with $L_{0}= 2.44\times 10^{-8}$~$\Omega$W/K).  This result confirms the purely electronic origin of $\kappa_{xy}$ and  conforms with reports for copper and the normal state of YBCO \cite{Ong}.

We next examine the scaling relation of $\kappa_{xy}$ with respect to $T$ and $H$ proposed in Ref. \cite{Simon}.  A scaling relation of the single variable $x=t/\sqrt{h}$ with $t=T/T_{c}$ and $h=H/H_{c2}^{orb} $ is derived as
\begin{equation}
\kappa_{xy}\sim T^2F_{\kappa_{xy}}(x),
\end{equation}
where $F(x)$ is a scaling function.  As shown in Fig.~3, $|\kappa_{xy}(T,H)|/T^2$ collapses into a common function of $x$ at $x\alt$ 0.07 at low temperatures within the error bar, suggesting a scaling relation, although not as prominent as in YBCO \cite{Zhang}.   The present scaling relation provides further support for $d$-wave symmetry in CeCoIn$_5$ \cite{Scaling}.

At first glance, the field dependence of $\kappa_{xx}$ does not look like what is expected for a nodal superconductor. In contrast to  fully gapped superconductors, heat transport in nodal superconductors is dominated by contributions from delocalized QP states rather than bound states associated with vortex cores \cite{Kubert,Barash,Vekhter,vekhter2,Franz}.  The most remarkable effect on the thermal transport is the Doppler shift of the QPs in the presence of supercurrents  around vortices. Usually, this effect leads to a $\sqrt{H}$ increase in the population of delocalized QPs and a subsequent increase in $\kappa_{xx} (H)$ that is nearly proportional to $\sqrt{H}$, as  experimentally observed in several unconventional superconductors.  The field dependence of $\kappa_{xx}$ observed for CeCoIn$_5$ does not show this behavior.  We will argue below, that this is a result of an increase of the DOS compensated by a reduction of the mean free path, both induced by the magnetic field.

The QP mean free path is directly provided by the thermal Hall angle in the weak field limit $\omega_c\tau\ll 1$,
\begin{equation}
 \tan \Theta \simeq \omega_c \tau \simeq \frac{e B \ell }{k_F \hbar},
 \end{equation}
where $\omega_c$ is the cyclotron frequency, $k_F$ is the Fermi wave number.  Figure 4 shows $|\tan \Theta|/B$ at the zero field limit  and at 5.2~T,  slightly above $H_{c2}$,  as a function of $T$, together with the electrical Hall angle $(\tan \Theta_e \equiv \sigma_{xy}/\sigma_{xx})$ divided by $B$.   The magnitude of $|\tan \Theta| / B$ coincides well with that of $|\tan \Theta_e|/B$ at the  zero field limit,  but at 5.2~T it is slightly larger.  Below the coherence temperature, $T^* \simeq$ 20~K shown by the arrow in Fig.~4, the resisitivity exhibits $T$-linear behavior.  Below $T^*$, the cotangent of the electrical Hall angle for $B\rightarrow 0$ was reported to display a $ T^2$ behavior, as shown by the dashed line,  which represents lim$_{B\rightarrow0}|\cot \Theta_e|/B=a+bT^2$ with $a$ = 4.38~T$^{-1}$ and $b$ = 0.20~K$^{-2}T^{-1}$ \cite{Nakajima}.  Below $T_c$, $|\tan \Theta|/ B$ increases much faster than the extrapolated temperature dependence observed above $T_c$. This enhancement of almost one order of magnitude is a direct evidence of a drastic increase in the QP mean free path
below $T_c$. The inset of Fig.~4 shows the  value of $\ell$ below $T_c$ using $k_F=1.85 \times 10^{9}$~cm$^{-1}$.  At $T$ = 0.46 K, $\ell$ has a value of 1.6~$\mu$m.

\begin{figure}[t]
\begin{center}
\includegraphics[height=60mm]{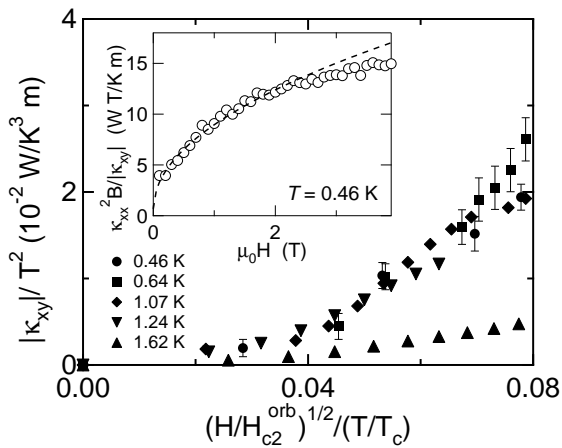}
\caption{Scaling plot $\kappa_{xy} / T^2 $ vs. $x=\sqrt{H/H_{c2}^{orb}} / (T/T_c)$. We used $H_{c2}^{orb}=$ 12~T.  The data collapse into the same curve at low temperatures. Inset: Field dependence of $\kappa_{xx}B/|\kappa_{xy}|$,  which is proportional to the DOS of the delocalized QPs $N_{del}(E)$ at 0.46~K. The dashed line indicates $\sqrt{H}$-dependence.  }
\end{center}
\end{figure}

\begin{figure}[t]
\begin{center}
\includegraphics[height=60mm]{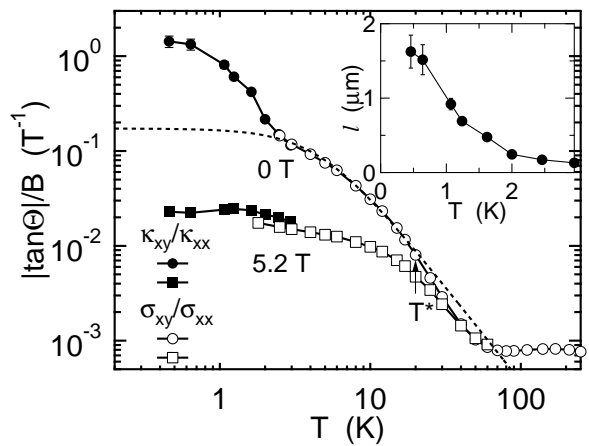}
\caption{Temperature dependence of  thermal Hall angle ($\bullet$  and $\blacksquare$), $\tan\Theta \equiv \kappa_{xy}/\kappa_{xx}$, and electrical Hall angle ($\circ$ and $\square$), $\tan \Theta_e \equiv \sigma_{xy}/\sigma_{xx}$,  divided by $B$ in zero field limit  and in the normal state at 5.2~T.   The dashed line represents the relation lim$_{B\rightarrow0}|\cot \Theta_e|/B=a+bT^2$.  $T^*$ is the coherence temperature. See the text for details.  Inset: Temperature dependence of the QP mean free path $\ell$ in zero field estimated from the thermal Hall angle (Eq.~(2)). }
\end{center}
\end{figure}

An alternative way of estimating the QP mean free path is to use the well-known link between $\kappa_{xx}$ and the specific heat $C_{e}$: $\kappa_{xx}=\frac{1}{3} C_{e} v_{F} \ell$. Now, at $T$~=~0.4K, with $\kappa_{xx}=0.48~W/K$ and $C_e$ = 0.056J/K$\cdot$ mol \cite{Movshovich}, if we take  $v_F$=2130 km/s (calculated using a Fermi energy, $\epsilon_{F}$, of 15K \cite{Kim} and a mass enhancement of $m^{*}=100 m_{e}$ \cite{Shishido}), the magnitude of $\ell$ of 1.1~$\mu$m is comparable to that  yielded by $\tan \Theta/ B$. This quantitative consistency also  confirms the very low value of $\epsilon_{F}$ deduced from the temperature dependence of specific heat \cite{Kim}.  

Figure 5 displays the field dependence of the QP mean free path at $T$ = 0.46~K.  As seen in the figure, the magnetic field dramatically suppresses the QP mean free path. Even at $H$ = 0.1~T ($H/H_{c2}^{orb} \alt $~1/100), $\ell$ is reduced by one order of magnitude. The inset of Fig.~5 shows the data on a log-log scale. For comparison, we plot the average distance between vortices $a_v=\sqrt{\Phi_0/B}$ by a dashed line.  At low fields, the QP mean free path is several times longer than the intervortex distance, but becomes comparable with  $a_v$ at higher fields.  This strong variation in the QP mean free path with magnetic field appears to be the origin of the unexpectedly flat field dependence of $\kappa_{xx}$ discussed above. In order to check whether the DOS of the {\it delocalized} QPs, $N_{del}(E)$, displays the expected $\sqrt{H}$ dependence, we can use the conjectures $\kappa_{xx} \propto N_{del}(E) \ell$ and $|\kappa_{xy}|/ (B \kappa_{xx}) \propto \ell$. Plotting $\kappa_{xx}^{^2}B/ |\kappa_{xy}|$ as a function of $H$  reveals the field dependence of $N_{del}(E)$. As seen in the inset of Fig.~3, this ratio displays a field-dependence close to the $\sqrt{H}$ behavior expected for a $d$-wave superconductor.  

The strong field dependence of $\ell$ is a feature that is not yet understood.  There are two lines of thought to understand the QP transport.   It has been argued that low energy QPs in a periodic vortex lattice are described by Bloch wavefunctions and are not scattered \cite{Kita}. In contrast, in a strongly disordered vortex lattice,  QP scattering is caused by Andreev scattering on the velocity field associated with the vortices.  In this case, the QP mean free path is proportional to $a_v$.  This argument was used to explain the ''plateau" in $\kappa_{xx}(H)$ observed in Bi2212, in which the vortex lattice is strongly distorted \cite{Krishana,Aubin,Franz}.  However, as indicated by small angle neutron scattering experiments \cite{Eskildsen}, there is no reason to assume that the vortex lattice in clean CeCoIn$_5$ is strongly distorted.

\begin{figure}[t]
\begin{center}
\includegraphics[height=60mm]{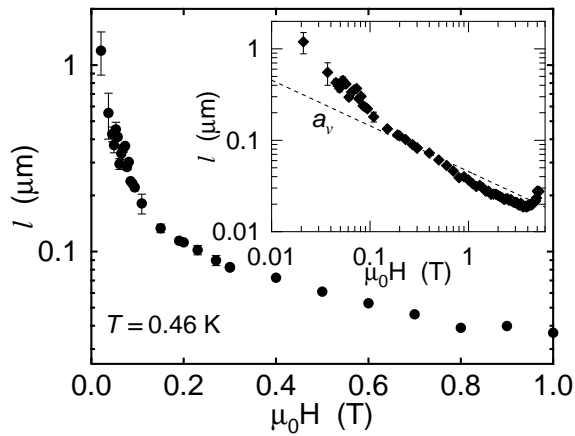}
\caption{Field dependence of the QP mean free path estimated from the thermal Hall angle below $H=$1~T at $T=$0.46~K.  Inset: The same data up to $H_{c2}$ plotted on a  log-log scale. The  dashed line represents the average vortex distance $a_v=\sqrt{\Phi_0/H}$. }
\end{center}
\end{figure}

The initial decrease of $\ell$ at low fields may be explained without invoking vortex scattering.  At very low fields, where the condition $\sqrt{H/H_{c2}^{orb}}<T/T_{c}$ is satisfied,  thermally excited QPs dominate over  Doppler shifted QPs. It has been shown that in this regime the DOS enhanced by the Doppler shift leads to a suppression of  the impurity scattering time \cite{Kubert,Barash,Vekhter,vekhter2}. It is yet to be seen if this  can explain the magnitude of the decrease observed at very low fields. It does not seem to be relevant to the $H$-dependence of $\ell$ at higher fields.

Although $\sqrt{H}$-dependent term in $\ell$ in a nearly periodic vortex lattice has been argued in Ref.\cite{vekhter2},  it is open question whether any deviation of the vortex from the perfect arrangement of the vortex lattice in  principle produces significant effects.    Several peculiarities of CeCoIn$_5$ may lead to unusual vortex-QP scattering.  In fact, the very strong suppression of $\ell$ up to $H_{c2}$ has never been observed in any other superconductors, including UPd$_2$Al$_3$ \cite{wata} and YNi$_2$B$_2$C \cite{izawa2} with similar $H_{c2}$ values.   One is the possible existence of antiferromagnetism in vortex cores. Several experiments indicate that the AFM phase is superseded by the superconducting transition \cite{QCP}.  This in turn suggests that the AFM correlation is strongly enhanced in the region around vortex cores \cite{Kakuyanagi}.  In this case the QPs may be significantly scattered by the AFM fluctuation in the core region.  Further investigation is strongly required to clarify the origin of the peculiar QP transport in CeCoIn$_5$.

Another feature to be considered is the energy scale of the QP spectrum in the vortex core set by the confinement energy $\hbar \omega_0 \sim \Delta^2/\varepsilon_F$.    For most superconductors, this energy level is negligibly small. For CeCoIn$_5$ however, $\varepsilon_F\sim$ 15 K and $\Delta\sim$~5~K, so that  $\hbar \omega_0\sim$~1.5~K and the vortex spectrum becomes important at low temperatures.  Moreover, when a vortex moves, energy dissipation is produced by the scattering of QPs within the vortex core.   If the broadening of the QP states ($\hbar/\tau$) turns out to be much smaller than the energy scale of the spectrum within the core, $\omega_0\tau\gg1$, the vortex system enters a new regime (the superclean regime) that is difficult to access in most superconductors.  The superclean condition is equivalent to $\ell/\xi \gg \varepsilon_F/\Delta$.  In CeCoIn$_5$,  $\ell\sim$1~$\mu$m and $\xi\sim$~5~nm yields $\ell/\xi\sim$~200 at low fields, which is much larger than $\varepsilon_F/\Delta\sim$~3.  This is in sharp contrast to other superconductors in which $\ell/\xi \ll \varepsilon_F/\Delta$.  In the superclean regime,  strong enhancement of the vortex viscosity, which leads to anomalous vortex dynamics including an extremely large vortex Hall angle, is expected \cite{Harris}.

To conclude, we have measured the thermal Hall angle of  CeCoIn$_5$ and found that it indicates a dramatic increase  in the quasiparticle mean free path below $T_c$.  In spite of the presence of a periodic vortex lattice, this enhancement is easily suppressed by a weak magnetic field.   These results highlight that  CeCoIn$_5$ is unique among superconductors.     We found that $\kappa_{xy}$ displays the scaling relation expected for $d$-wave symmetry.  Moreover the DOS of the delocalized quasiparticles obtained from the thermal Hall conductivity,  are consistent with $d$-wave symmetry.  Finally, the results indicate that CeCoIn$_5$ is in the superclean regime.

We thank S.~Fujimoto, R.~Ikeda, Y.~Kato, T.~Kita, H.~Kontani, and I.~Vekhter for their helpful discussions.


\begin{thebibliography}{99}
\bibitem{measure} Below 0.4~K  $\kappa_{xy}$ can not be extracted accurately in the  low field region due to a  finite hysteresis arising from the vortex pinning effect.
\bibitem{Scaling} An additional scaling relation $\kappa_{xy}\sim T\sqrt{H}$ reported in YBCO \cite{Zhang} was not observed here .
\end{thebibliography}
\end{document}